# Light Induced Surface Tension Gradients for Hierarchical Assembly of Particles from Liquid Metals

Jiayun Liang[1], Zakaria Y. Al Balushi[1,2]*

[1]Department of Materials Science and Engineering, University of California, Berkeley, Berkeley, CA 94720, USA.

[2]Materials Sciences Division, Lawrence Berkeley National Laboratory, Berkeley, CA 94720, USA.

**ABSTRACT**

Achieving control over the motion of dissolved particles in liquid metals is of importance for the meticulous realization of hierarchical particle assemblies in a variety of nanofabrication processes. Brownian forces can impede the motion of such particles, impacting the degree of perfection that can be realized in assembled structures. Here we show that light induced Marangoni flow in liquid metals (i.e., liquid-gallium) with Laguerre-gaussian ($LG_{pl}$) lasers as heating sources, is an effective approach to overcome Brownian forces on particles, giving rise to predictable assemblies with high degree of order. We show that by carefully engineering surface tension gradients in liquid-gallium using non-gaussian $LG_{pl}$ lasers, the Marangoni and convective flow that develops in the fluid drives the trajectory of randomly dispersed particles to assemble into $100\ \mu m$ wide ring-shaped particle assemblies. Careful control over the parameters of the $LG_{pl}$ laser (i.e., laser mode, spot size, and intensity of the electric field) can tune the temperature and fluid dynamics of the liquid-gallium as well as the balance of forces on the particle. This in turn can tune the structure of the ring-shaped particle assembly with a high degree of fidelity. The use of light to control the motion of particles in liquid metals represents a tunable and rapidly reconfigurable approach to spatially design surface tension gradients in fluids for more complex assembly of particles and small-scale solutes. This work can be extended to a variety of liquid-metals, complementary to what has been realized in particle assembly out of ferrofluids using magnetic fields.

*Corresponding author, e-mail: albalushi@berkeley.edu

Achieving control over the motion of particles in fluids is important to realize the high degree of perfection needed in the assembly processes of nano-to-macroscale objects [1]. An important force that must be taken into consideration in the assembly of such particles is Brownian motion [2]. One approach to reduce the impact of Brownian on particles dispersed in liquids is by taking advantage of the Marangoni effect [3]. Marangoni effect (i.e., Marangoni flow) describes fluid flow induced by surface tension gradients. Liquids with higher local surface tension pulls strongly around its surrounding. The resulting shear stress at the two-phase interface gives rise to a rather strong convective motion of the fluid, naturally flowing from regions of low local surface tension to high ones. This surface tension gradient in fluids can be induced by several means, for example by adding surface surfactants [4,5], tailoring the solute concentration [6] or developing a temperature gradient on the surface using a variety of heating sources [7]. Much of the research in this field has been focused on controlling Marangoni flow in transparent hemispherical droplets. In this case, fluid flows away from the apex of the droplet to compensate for the liquid loss at the edges. Particles dispersed in these droplets end up accumulating at the edges, giving rise to the seminal "coffee-ring" effect [8]. By inducing Marangoni flow in the opposite direction, one can suppress the "coffee-ring" effect and achieve a more uniform particle pattern [4], or even the "reverse coffee-ring" effect [9], where particles accumulate at the center of the droplet upon evaporation. Furthermore, the Marangoni effect has also been observed in opaque liquids, like liquid metals, such as in a variety of crystal growth directional solidification processes [10]. Such liquid metals have also been used as catalysts during the fabrication of thin films and nanostructures by the vapor-liquid-solid (VLS, [11,12,13]), liquid-liquid-solid (LLS, [14]), and solid-liquid-solid (SLS, [15]) growth mechanisms of materials. The limitation of these crystal growth processes was that the size and geometry of the precipitating crystalline solid out of the liquid metal is always confined by the geometry of the liquid metal itself [13,16,17]. Therefore, if one could control the fluid flow of opaque liquids and therefore the motion of small-scale solutes during the fabrication of materials or assemblies out of liquid metals, it could be possible to create material designs out of the liquid metal with varying degrees of complexity.

Herein, we model the light-matter interaction of laser heating sources to spatially structure surface tension gradients on liquid metals to therefore control the motion and assembly of particles *via* the Marangoni effect. In this work, Laguerre-Gaussian ($LG_{pl}$) lasers were utilized to engineer surface tension gradients and induced Marangoni flow. Here, $p$ and $l$ are radial and azimuthal index numbers for the $LG_{pl}$ laser, respectively. The $LG_{pl}$ lasers were selected because compared to a Gaussian laser, it operates in higher-order transverse modes which determines the intensity distribution in the cross-section of the laser beam, where higher-order transverse modes lead to a higher complexity in this intensity distribution. Therefore, $LG_{pl}$ lasers allow for more degrees of freedom in the design of the temperature gradient and thus Marangoni flow patterns. In this work, liquid-gallium was used as a model liquid metal because of its low melting point, high boiling point and relatively low vapor pressure over a wide temperature range [18]. Furthermore, liquid-gallium does not form compounds with many elements (e.g., tungsten, silicon, germanium, carbon etc., Table S1) and therefore useful in a variety of crystal growth processes [12,17,19]. In this work, a ring pattern assembly of tungsten particles out of liquid-gallium was realized by strategically choosing the radial and azimuthal index numbers as well as beam conditions for the $LG_{pl}$ laser interacting with the surface of liquid-gallium. The assembled ring-shaped particle patterns were preferentially formed because of unique fluid flow vortices that form within the bulk of liquid-gallium, essentially creating a "forbidden zone" of assembly. This allows one to specifically structure assembled particle features between the "coffee-ring" and "reverse coffee-ring" effects. The perfection of the assembly was quantified by assessing the entropy of these patterns, which improved with the existence of significant drag forces on the particles that could overcome Brownian motion of the particles due to Marangoni flow.

## RESULTS AND DISCUSSION

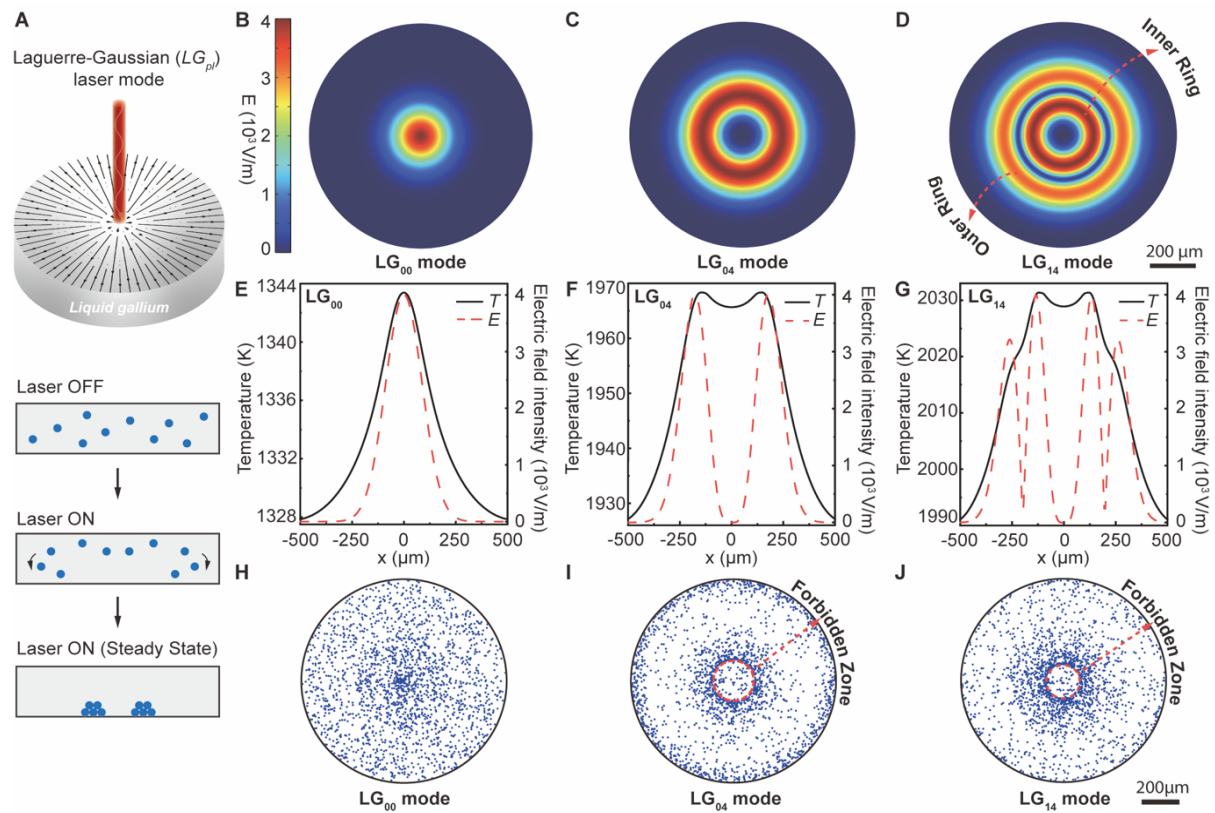

Figure 1. Electric field profiles, temperature profiles and resulting particle patterns assembled out of liquid-gallium under different Laguerre-Gaussian ($LG_{pl}$) laser modes. (A) Schematic highlighting the process of assembling particles out of liquid-gallium using a $LG_{pl}$ lasers (laser wavelength, $\lambda = 645\ nm$ and spot size, $w_o = 125\ \mu m$). When $LG_{pl}$ lasers interact with liquid-gallium, Marangoni flow develops, guiding randomly dispersed tungsten particles (diameter, $d_p = 20\ \mu m$) to assemble at the liquid-solid interface. (B, C and D) Profile of the electric field on the surface of liquid-gallium under (B) $LG_{00}$, (C) $LG_{04}$, and (D) $LG_{14}$ lasers, respectively. An inner and outer ring are observed in the doughnut-shaped electric field profile for (D) $LG_{14}$ laser, as highlighted by two *dashed red arrows*. The scale bar for all the electric field profiles is $200\ \mu m$. (E, F and G) Profiles of the electric field (*dashed red lines*) and temperature (*solid black lines*) on the surface of liquid-gallium. (H, I and J) Resulting particle patterns at the liquid-solid interface for (H) $LG_{00}$, (I) $LG_{04}$, and (J) $LG_{14}$ lasers interacting with liquid-gallium. The ring-shaped particle patterns are obtained with (I) $LG_{04}$, and (J) $LG_{14}$ laser modes only, creating a "forbidden zone" of assembly, where tungsten particles do not gather which is highlighted by *dashed red lines*. The scale bar for all particle patterns is $200\ \mu m$.

Laser heating is useful in engineering fluid flow [20]. Lasers can also be tuned through a variety of parameters, including wavelength ($\lambda$), spot size ($w_o$) and the profile of the electric field. This makes lasers highly accessible as rapid and reconfigurable heating sources for engineering temperature gradients in fluids. So far, Gaussian lasers have only been applied as a heating source to induce Marangoni flow at liquid-solid and liquid-liquid interfaces of transparent liquids [21]. The use of non-Gaussian lasers to engineer more complex temperature gradients in transparent or opaque liquids has been largely unexplored. Therefore, to illustrate the uniqueness of using a laser to structure the surface tension gradient and therefore convective flow patterns in liquid metals, the interaction of $LG_{pl}$ lasers ($\lambda = 645\ nm$ and $w_o = 125\ \mu m$, Figure 1) onto the surface of liquid-gallium with 2000 randomly dispersed tungsten particles with diameter ($d_p$) of 20 $\mu m$ was investigated. As illustrated in Figure 1A, the interaction of the laser will induce Marangoni flow for the hierarchical assembly of these randomly dispersed particles at the liquid-solid interface.

To reveal the relationship between the $LG_{pl}$ laser and the Marangoni flow, three different $LG_{pl}$ lasers (i.e., $LG_{00}$, $LG_{04}$, and $LG_{14}$, see **Methods**) were investigated. The electric field distribution ($E$) on the surface of the liquid-gallium is shown in Figure 1 (B, C, and D). Unlike the Gaussian distribution of the electric field with the $LG_{00}$ laser (Figure 1B), the electric field distribution of the $LG_{04}$ and $LG_{14}$ lasers were doughnut-shaped (Figure 1, C and D, respectively). That is, the peak intensity of the electric field was not at the center of the surface of liquid-gallium (Figure 1, E, F, and G, *dashed red lines*). Unlike the single electric field ring maximum in the $LG_{04}$ laser seen in Figure 1C, the $LG_{14}$ laser contained two ring maxima in the electric field distribution over the surface of liquid-gallium. These rings are labeled as "inner" and "outer" rings in Figure 1D (*dashed red arrows*). The interaction of these lasers with liquid-gallium leads to electromagnetic heating ($Q_e$) of the surface, which is given by:

$$Q_e = \frac{1}{2} Re(\vec{J} \cdot \vec{E^*}) = \frac{1}{2} Re(\sigma_f \vec{E} \cdot \vec{E^*})$$

where $\sigma_f$ is the electrical conductivity and $J$ is the current density of liquid-gallium. The surface temperature profiles of liquid-gallium due to the interaction of these three $LG_{pl}$ lasers are also illustrated in Figure 1 (E, F, and G). Since the intensity of the electromagnetic heating was proportional to the square of the strength of the electric field, the surface temperature profiles of liquid-gallium were composed of single (Figure 1E), double (Figure 1F), and quad (Figure 1G) temperature maxima for the $LG_{00}$, $LG_{04}$, and $LG_{14}$ lasers, respectively. Though the maximum intensity values of the electric field were equal for the $LG_{00}$, $LG_{04}$, and $LG_{14}$ lasers ($4 \times 10^3 V/m$), due to the variation in the electric field distribution between these three lasers, the amount of electromagnetic heating and therefore resulting temperature maxima of liquid-gallium using each laser differed. This strong dependance of the electromagnetic heating of liquid-gallium to the electric field of the laser highlights the ability to engineer a variety of unique temperature gradients in liquid-gallium with a high degree of versatility.

Moreover, the temperature gradient that develops on the surface of liquid-gallium induces sheer stresses at the gas-liquid interface giving rise to Marangoni flow:

$$\left( -P\vec{I} + \mu(\nabla\vec{v} + (\nabla\vec{v})^T) - \frac{2}{3}\mu(\nabla \cdot \vec{v})\vec{I} \right)\vec{n} = k_0 \nabla_t T$$

where $\vec{v}$ is the velocity of the fluid and $k_0$ is the temperature coefficient of the surface tension [22]. Over time, the randomly dispersed tungsten particles pick up drag forces and follow the convective flow streamlines within the bulk of the liquid-gallium. Finally, these particles assemble at the liquid-solid interface into a variety of patterns which depended on the conditions of the $LG_{pl}$ laser. The resulting particle patterns are highlighted in Figure 1 (H, I, and J). Figure 1H shows randomly distributed tungsten particles across the solid surface when the $LG_{00}$ laser was used as a heating source. However, when the $LG_{04}$, and $LG_{14}$ lasers were utilized, ring-shaped particle patterns were assembled out of the liquid-solid interface as shown in Figure 1I and 1J for the $LG_{04}$, and $LG_{14}$ lasers, respectively. The use of the $LG_{04}$, and $LG_{14}$ lasers essentially created regions where particles could not assemble on the solid surface. These regions are referred to as the "forbidden zone" which form due to unique vortices that develop within the flow pattern in the bulk of liquid-gallium.

**Influence of Marangoni Flow on Brownian and Particle Assembly:**

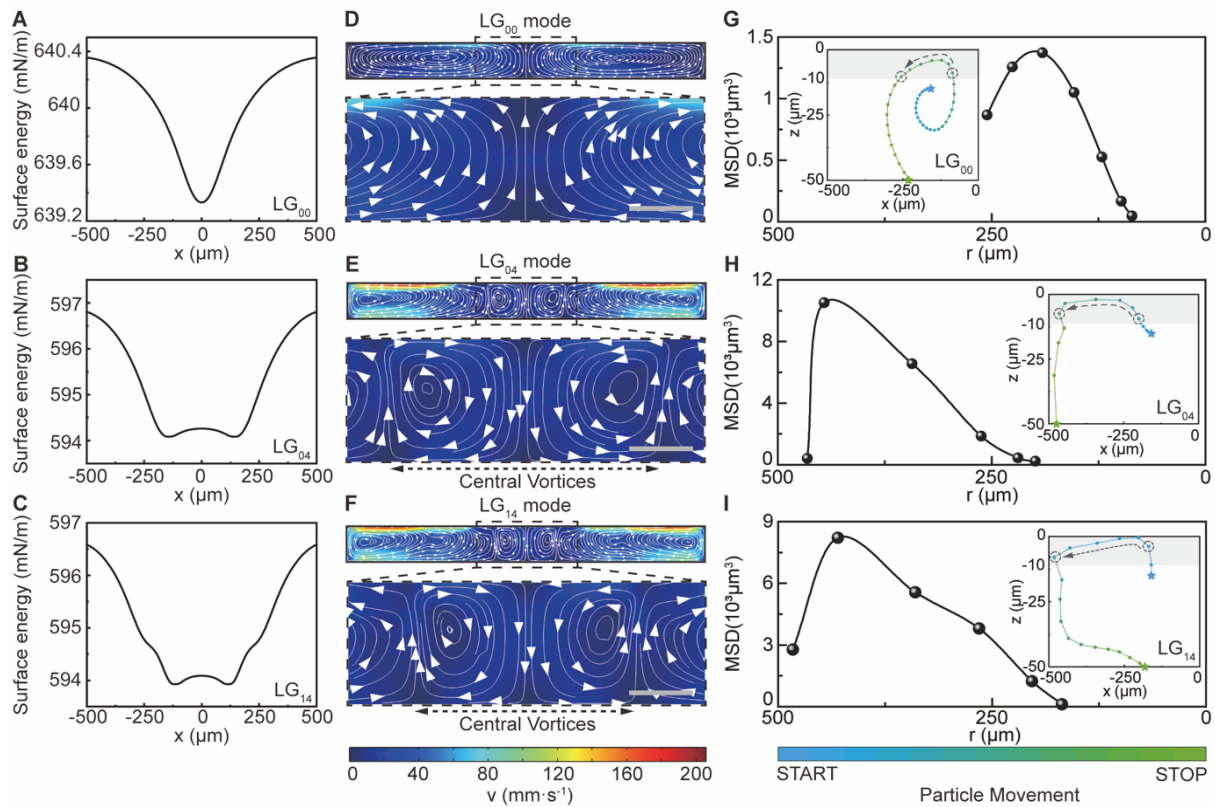

**Figure 2. Influence of surface tension gradients of liquid-gallium on the particle motion in the bulk fluid. (A, B, and C)** Surface tension profiles of liquid-gallium along the x-axis for **(A)** $LG_{00}$, **(B)** $LG_{04}$, and **(C)** $LG_{14}$ lasers ($\lambda = 645\ nm$ and $w_o = 125\ \mu m$). **(D, E, and F)** Convective flow patterns induced by the surface tension gradient (i.e., Marangoni flow) for **(D)** $LG_{00}$, **(E)** $LG_{04}$, and **(F)** $LG_{14}$ lasers. For **(E)** $LG_{04}$, and **(F)** $LG_{14}$ lasers, two additional central vortices are observed in the convective flow patterns in the fluid, as highlighted by *dashed black lines* under the magnified regions (scale bars are 50 $\mu m$). The absolute value of the fluid velocity is indicated by the color bar at the bottom. **(G, H, and I)** The mean-squared displacement (MSD) versus radial position ($r$) of for particles under **(G)** $LG_{00}$, **(H)** $LG_{04}$, and **(I)** $LG_{14}$ lasers when the particle moves near the top surface of liquid-gallium ($|z| \leq 10\ \mu m$, *shaded gray background*). The insets in (G, H, and I) are the particle trajectories (*blue-to-green gradient*) guided by the fluid flow induced under the **(G)** $LG_{00}$, **(H)** $LG_{04}$, and **(I)** $LG_{14}$ laser modes, respectively. All particles were released at the same position in the fluid (*blue star*) and finally assembled at the liquid-solid interface (*green star*).

The pattern of the assembled particles observed in Figure 1 (H, I, and J) were highly dependent on the electric field distribution and the consequentially developed surface tension gradients when the $LG_{00}$ (Figure 2A), $LG_{04}$ (Figure 2B), and $LG_{14}$ (Figure 2C) lasers interacted with the surface of liquid-gallium. To understand the origin of the "forbidden zone", the Marangoni and resulting convective flow patterns within the depth of liquid-gallium, as a function of the laser, were investigated (Figure 2, D, E, and F). Fluid flow induced by the $LG_{00}$ laser (Figure 2D) led to the formation of two vortices, allowing liquid-gallium to recirculate from the center to the periphery of the bulk. However, the formation of two additional vortices within the center of the convective flow pattern (i.e., "central vortices") were prominent when the $LG_{04}$ (Figure 2E) and $LG_{14}$ (Figure 2F) lasers interacted with liquid-gallium. A substantial enhancement in the velocity of the fluid can also be seen in Figure 2E and 2F. This enhanced fluid velocity will strongly dictate the trajectory of the particles dispersed in the fluid. This is evident in Figure 2 (G, H, and I), which highlights the mean-squared displacement (MSD, see **Supporting Information – Sec. II**), for particles moving near the surface of liquid-gallium ($|z| \leq 10 \ \mu m$) as a function of the $LG_{pl}$ laser. The corresponding $xz$ trajectory of the particles are also included as insets in Figure 2 (G, H, and I). In the case of the $LG_{04}$ (Figure 2H) and $LG_{14}$ (Figure 2I) lasers, the peak of the MSD shifts further away from the center of liquid-gallium compared to the $LG_{00}$ laser (Figure 2G). Since the values of the MSD reflects the dynamics of the surrounding fluid, the emergence of these new "central vortices" with outward flow will ultimately pay a key role in preventing particles from remaining at the center of the fluid.

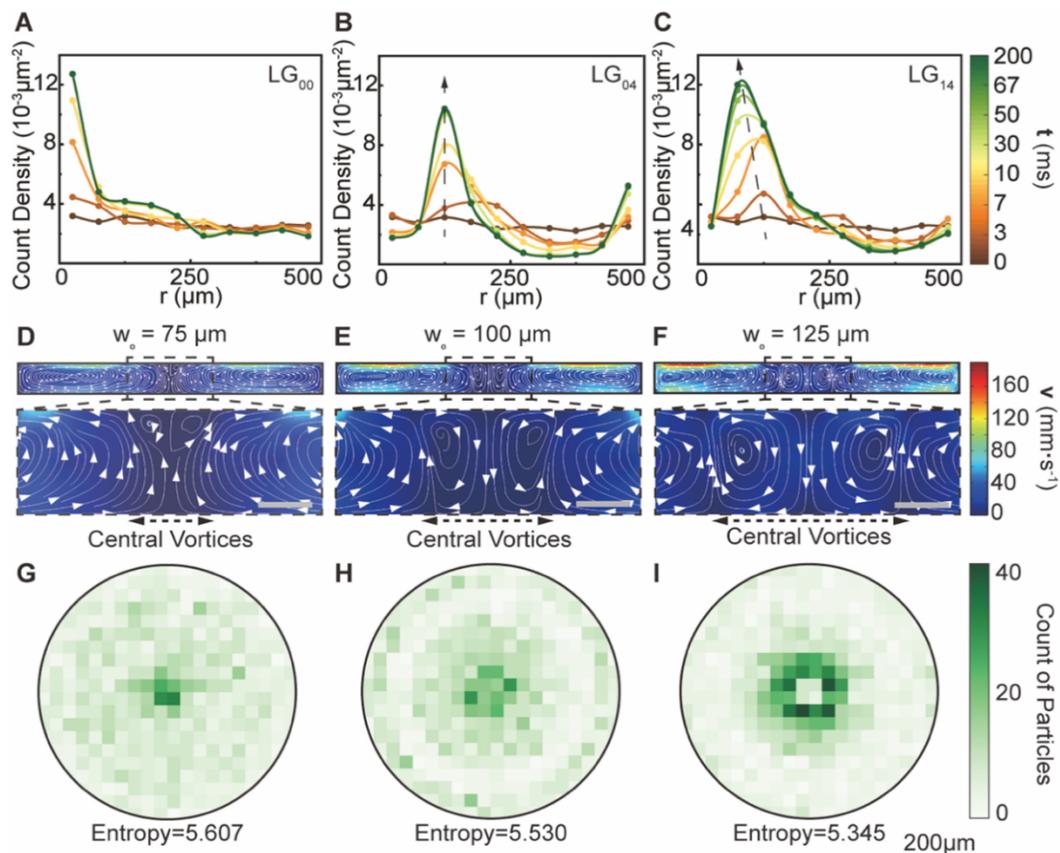

**Figure 3. Impact of the laser mode and spot size on the particle pattern assembled out of liquid-gallium. (A, B, and C)** Time dependent radial ($r$) particle count density for 2000 randomly released tungsten particles ($d_p = 20 \ \mu m$) in liquid-gallium guided by the convective flow induced by (A) $LG_{00}$, (B) $LG_{04}$, and (C) $LG_{14}$ lasers. All 2000 particles were released at $t = 0 \ s$ and all settled at the liquid-solid interface at $t = 0.2 \ s$. The width of the radial bin for the particle count density profiles was set to $50 \ \mu m$. **(D, E, and F)** Fluid flow pattern in liquid-gallium under $LG_{14}$ laser with the spot size of (A) $75 \ \mu m$, (C) $100 \ \mu m$, and (E) $125 \ \mu m$. Two additional "central vortices" are magnified for clarity (scale bar are $50 \ \mu m$). The absolute value of the fluid velocity is indicated by the color bar (*red-to-blue gradient*) on the right. **(G, H, and I)** Density distribution maps (bin size = $50 \ \mu m$) and corresponding entropy values for particle patterns induced by

fluid flow under the $LG_{14}$ laser with a spot size ($w_o$) of (G) 75 $\mu m$, (H) 100 $\mu m$, and (I) 125 $\mu m$. The scale bar for all density distribution maps is 200 $\mu m$.

The spatial extent of the "central vortices", and therefore the "forbidden zone", was smaller in the $LG_{14}$ laser compared to that in the $LG_{04}$ laser. This was due to the existence of the "inner" and "outer" rings in the electric field distribution as alluded to in Figure 1D. More broadly, the influence of the "central vortices" on the spatial extent of the "forbidden zone" can be further supported by assessing the time dependence of the radial ($r$) particle count density of 2000 randomly released tungsten particles ($d_p = 20\ \mu m$, Figure 3, A, B, and C). When particles were initially released ($t = 0s$), the particle count density across the radial direction of the liquid-solid interface were very close to each other, oscillating around $3.15 \times 10^{-3}\ \mu m^{-2}$ with a mean-squared error of $0.27 \times 10^{-3}\ \mu m^{-2}$. Over time, the particle count density at $r \leq 50\ \mu m$ in $LG_{00}$ increased (Figure 3A), while in $LG_{04}$ and $LG_{14}$ it decreased (Figure 3, B and C). At $t = 0.2\ s$, the particle count density for the $LG_{14}$ laser around the central region of the liquid-solid interface was $2.55 \times 10^{-3}\ \mu m^{-2}$ (Figure 3C, *solid green line*), which was an 80% reduction in the particle count density when compared to the $LG_{00}$ laser ($12.73 \times 10^{-3}\ \mu m^{-2}$, Figure 3A, *solid green line*). Also, as time increased, the peaks in the particle count density also increased (Figure 3, B and C, *dashed black arrows*). The peak positions for the $LG_{04}$ and $LG_{14}$ lasers, were at $r = 125\ \mu m$ and $75\ \mu m$ from the center of the liquid-solid interface, respectively. This elucidates why a smaller diameter in the "forbidden zone" was observed in the $LG_{14}$ laser when compared to $LG_{04}$ laser (Figure 1, I and J, respectively). Therefore, the existence of ring features in the electric field distribution in the $LG_{pl}$ laser allows one to spatially control the extent of the central vortices, and by controlling other parameters of the laser beam itself could further tune the extent of the "forbidden zone".

Moreover, the significance of the central vortices on particle assembly at the liquid-solid interface was further corroborated by investigating the effect of the $LG_{14}$ laser spot size ($w_o$) on fluid flow. Evident in Figure 3D was the fluid flow that results from the interaction of a $75\ \mu m\ LG_{14}$ laser with liquid-gallium. At this spot size, the spatial extent of the central vortices in the fluid was rather negligible. However, as the $w_o$ increased to $100\ \mu m$ (Figure 3E) and $125\ \mu m$ (Figure 3F), the appearance of the "central vortices" was substantial, which increased in extent with increasing $w_o$. The absence of a fully developed central vortex in Figure 3D allowed for particles to freely circulate at the center of the liquid. In this case, particles that would assemble out of the liquid-solid interface would have a lower degree of order. To highlight this, we introduce the concept of entropy (see **Supporting Information – Sec. III**) to quantify this degree of order from the particle density maps as a function of $w_o$ (Figure 3, G, H, and I). A decrease in the value of entropy alludes to a higher degree of order in the particle distribution maps. As the $w_o$ increased from 75 $\mu m$, to 100 $\mu m$ and to 125 $\mu m$, the value of entropy was reduced from 5.607, 5.530 and 5.345, respectively. This means that the 125 $\mu m\ w_o$ for the $LG_{14}$ laser had the highest degree of order as reflected in the well-defined ring feature of the assembled particles seen in the distribution map in Figure 3F. Interestingly, it can be observed that 54.2% of the particles preferred to gather around the "forbidden zone" of the solid surface. This was because as the $w_o$ increased, the fully developed central vortices repelled particles from entering the central region of the fluid. This shows that the drag forces on the particle can significantly overcome Brownian forces to assemble patterns with a high degree of order.

**Investigating the Interacting Forces Impacting Particle Assembly:**

To gain a deeper understanding on how particles gather around the periphery of the "forbidden zone" when the $LG_{14}$ laser ($w_o = 125\ \mu m$) was utilized, the interacting forces on the moving particles in the fluid were investigated. Evidently, the synergy between the Marangoni flow and the effective gravitational forces on the particle was responsible in driving the formation of the ring-shaped patterns. This effective gravitational force ($G_{eff}$) is given by:

$$G_{eff} = G_p - F_{buoyant} = \frac{\pi}{6}(\rho_p - \rho_f)g d_p^3$$

where $G_p$, $F_{buoyant}$ and $d_p$ are the gravitational force, the buoyant force, and the diameter of the particle, respectively. Since the density of the tungsten particle ($\rho_p$) was higher than that of the fluid ($\rho_f$), the direction of the $G_{eff}$ points vertically downwards. The drag force ($F_D$) from the Marangoni flow and/or convective flow around the particle is given by:

$$\vec{F_D} = \frac{m_P(\vec{u} - \vec{v})}{\tau_P}$$

where $\vec{u}$ is the velocity of particle, $\vec{v}$ is the velocity of fluid flow. The characteristic time for the particle ($\tau_P$) is expressed by:

$$\tau_P = \frac{4\rho_p d_p^2}{3\mu C_D Re_r}$$

where $\mu$ is the dynamic viscosity of liquid gallium and $Re_r$ is the relative Reynolds number of particles in the flow. $C_D$ is the drag coefficient coupling the velocity of a particle to the surrounding fluid velocity. Relative Reynolds number ($Re_r$) is calculated with:

$$Re_r = \frac{|\vec{u} - \vec{v}| d_p \rho_f}{\mu}$$

And the relationship between $C_D$ and $Re_p$ is:

$$\begin{cases} C_D = \frac{24}{Re_r}, & Re_r < 1\ (Stokes\ model) \\ C_D = \frac{24}{Re_r}(1 + 0.15 Re_r^{0.687}) & 1 < Re_r < 1000\ (Schiller-Naumann\ model) \end{cases}$$

Considering the $Re_r$ is less than 1 for most of the time steps in our simulations, if not otherwise specified, results shown in this work is simulated with drag coefficient derived from Stokes law ($C_D = \frac{24}{Re_r}$).

The motion of a representative particle moving through the fluid can be seen in Figure 4A with magnified snapshots of this motion with the interacting $G_{eff}$ and $F_D$ forces superimposed in Figure 4 (B, C, and D). In Figure 4A, the particle picks up drag forces near the surface due to the Marangoni flow. However, before it reaches the sidewalls (Figure 4B), it moves downwards due to the gravitational force. The particle then continues to be guided by the convective flow of the fluid while slowly nearing the bottom of the liquid-solid interface (Figure 4C). The velocity of the fluid approaches zero near the boundary of the "forbidden zone" because the sum of fluid flow directions cancels each other out (Figure 4D). Hence, when particles approach this boundary, the gravitational force on the particle dominate its trajectory. As a result, the particle slowly settled around the periphery of the "forbidden zone", forming the well-defined ring-shaped pattern seen in Figure 3I.

To further elucidate the role of the gravitational force on the particle pattern formation, a comparative study on the motion of the particles without effective gravitational forces ($G_{eff} = 0$) was performed. As seen in Figure 4E, when a particle initially picked up drag forces due to the Marangoni flow (Figure 4F), it continued to be guided by the fluid flow near the surface until it nears the sidewalls of the liquid-gallium (Figure 4G). There the particle has a much higher probability to strike the sidewall because of the absence of gravitational forces. Due to the boundary conditions set in this model, this particle will ultimately bounce off the sidewall back into the fluid (See **Methods**). The particle then continues to be guided by the recirculating convective flow pattern of the fluid (Figure 4H), finally settling at the liquid-solid interface around the sidewall boundaries. Moreover, the influence of the sidewall boundary conditions on the particle motion and particle pattern formation at the liquid-solid interface was also investigated. Despite setting the sticking coefficient of the sidewalls to unity (i.e., all particles that strike the sidewall will stick), remanence of a ring-shaped particle pattern was still observed for particles with diameter of $20 \ \mu m$ (Figure S1A). More importantly, when the particle diameter was reduced to $5 \ \mu m$, less particles struck the sidewalls, giving rise to a more well-defined ring-shaped pattern (Figure S1B). This was because the probability of impacting the side wall was reduced with decreasing particle size (more information in **Supporting Information – Sec. IV**). Therefore, these results clearly demonstrate that the gravitational force on the particles reduces the particle-sidewall interaction, promoting the formation of a ring-shaped pattern assembly, where the degree of order for the pattern can be further improved through careful tuning of the parameters of the laser itself.

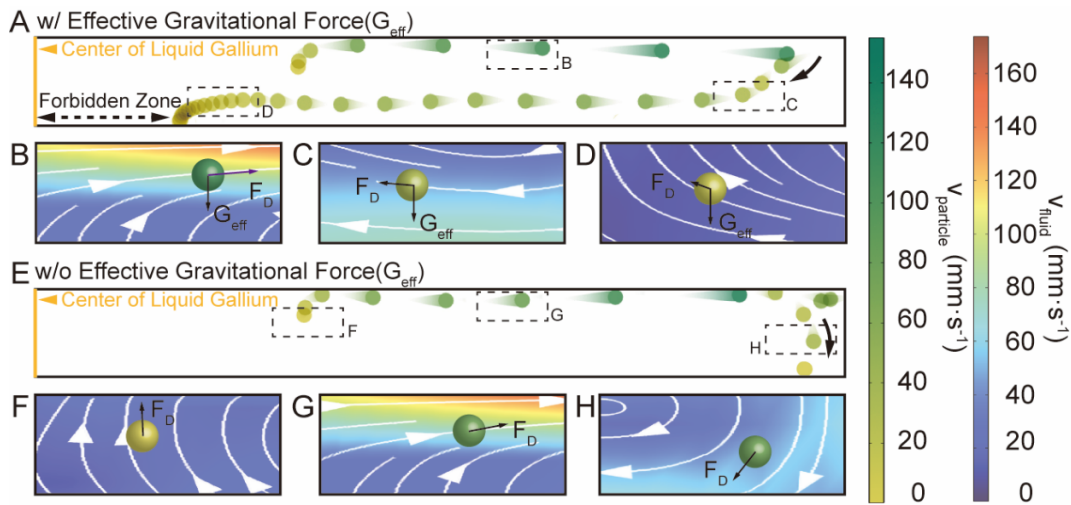

**Figure 4. Mechanism for the formation of ring-shaped particle patterns in liquid-gallium under the $LG_{14}$ laser. (A) Trajectory of a representative particle with effective gravitational force. The particle starts near the top surface and settles at the liquid-solid interface (bottom surface). For simplification, only half of the liquid-gallium fluid flow is shown, where the center of liquid-gallium is marked by a *solid yellow line*. The width of the "forbidden zone" is marked by a *double headed dashed black arrow*. (B, C, and D) Highlight the forces exserted on the particle when it: (B) moves along the top surface of liquid-gallium, (C) moves towards the liquid-solid interface and (D) before it settles at the liquid-solid interface. The direction of the drag forces ($F_D$) and the effective gravitational force ($G_{eff}$) are shown on the particle (*solid black arrows*). The background of the fluid velocity with superimposed streamlines illustrates the fluid dynamics**

of liquid-gallium in the magnified regions. (E) Trajectory of a representative particle without effective gravitational force. Particle starts near the top surface and settles near the periphery of the liquid-gallium boundary. (F, G, and H) Highlight the forces exserted on the particle when it: (F) initially is released into the fluid, (G) moves along the top surface of liquid-gallium, and (H) before it settles around the periphery of the liquid-gallium boundary. The absolute values of the fluid velocity (*red-to-blue gradient*) and particle velocity (*green-to-yellow gradient*) are indicated by the color bars on the right side of the figure.

**Tunable Laser Parameters for Particle Assembly:**

Our results showed that $LG_{pl}$ lasers can be effectively used to engineer the temperature and surface tension of liquid metals to create unique Marangoni flow patterns in the fluid. Although only a ring-shaped pattern assembly was demonstrated in this study, the use of non-Gaussian lasers can more broadly be a simple yet powerful approach to realize hierarchical assembly of particles and other small-scale solutes from liquid metals through appropriate scaling and laser parameter tuning. Therefore, in Figure 5, A to D, we summarize what are deemed the most important and tunable parameters of $LG_{pl}$ lasers to design ring-shaped particle assemblies with varying degrees of order. In Figure 5A, the influence of the $LG_{pl}$ lasers on the entropy of the ring-shaped pattern are highlighted. These results show that regardless of the bin size applied to calculate the particle density maps per $LG_{pl}$ laser (Figure 5A, inset), the $LG_{14}$ laser led to particle assemblies with the lowest entropy values when compared to the $LG_{00}$ and $LG_{04}$ lasers. Also, as evident in Figure 5B, enlarging the $w_o$ of the $LG_{14}$ laser mode was beneficial in improving the degree of order in the ring-shaped pattern assembly independent of the bin size used to calculate the entropy.

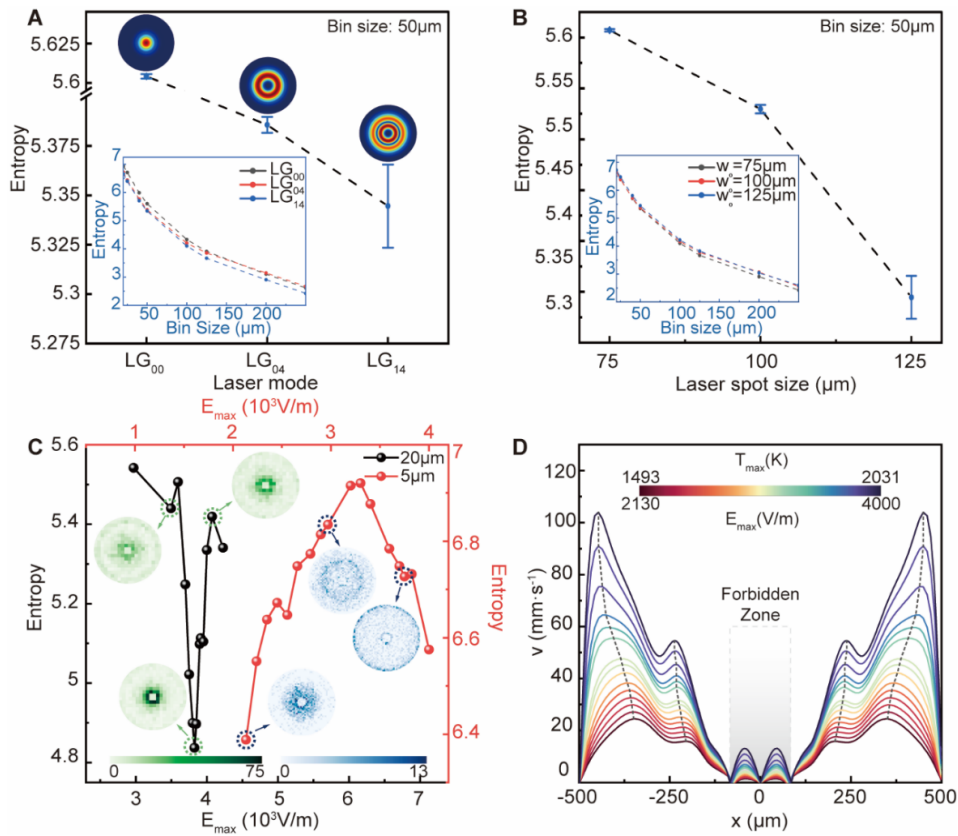

**Figure 5. Summary of the tuning knobs of $LG_{pl}$ lasers to obtain ring-shaped pattern with different degrees of order.** (A) Impact of $LG_{pl}$ lasers: entropy of particle patterns ($d_p = 20\ \mu m$) under different $LG_{pl}$ lasers ($LG_{00}$, $LG_{04}$, and $LG_{14}$, $w_0 = 125\ \mu m$, and $E_{max} = 4000\ V/m$). Bin size for the calculation of entropy is $50\ \mu m$. The entropy as a function of the bin sizes is also shown in the inset. **(B)** Impact of $LG_{pl}$ laser spot size: entropy of particle patterns ($d_p = 20\ \mu m$) under $LG_{14}$ laser ($E_{max} = 4000\ V/m$) with spot size of $75\ \mu m$, $100\ \mu m$, and $125\ \mu m$. Bin size for the calculation of entropy is $50\ \mu m$. The entropy values at different bin sizes are also shown in the inset. **(C)** Impact of maximum intensity value of the electric field: entropy of particle patterns under different $E_{max}$ values ($LG_{14}$ laser and $w_0 = 125\ \mu m$). The bin size for the calculation of entropy is $50\ \mu m$ for $20\ \mu m$ diameter particles (*solid black line*), and the bin size is $20\ \mu m$ for $5\ \mu m$ diameter particles (*solid red line*). Corresponding density distribution maps for selected data points (*dashed circles* in (C)) are shown. The count of particles in each square grid is indicated by the color bar at the bottom (*green-to-white gradient* for $20\ \mu m$ diameter particles, and *blue-to-white gradient* for $5\ \mu m$ diameter particles). **(D)** Fluid velocity profile along the solid-liquid interface under $E_{max}$ values. The velocity peaks are marked by *dashed black lines*. The region of the "forbidden zone" is highlighted in a *shaded grey region*. $E_{max}$ and temperature maxima ($T_{max}$) in the liquid-gallium are indicated by the top color bar.

Another important parameter that can be controlled in the laser is the maximum intensity value of the electric field ($E_{max}$). This allows one to tune the electromagnetic heating of the surface of liquid-gallium. This is highlighted in Figure 5C for both $20\ \mu m$ (*solid black line*) and $5\ \mu m$ (*solid red line*) particle sizes that assemble out of the liquid-gallium using a $LG_{14}$ laser. The resulting particle density maps that are shown as insets in Figure 5C contained the same number of particles for each electric field condition. For the case of the $20\ \mu m$ particle size, the entropy values were larger at lower values of the electric field. This was due to the reduced velocity and temperature of the fluid, and therefore drag forces on these $20\ \mu m$ particles. Interestingly, when increasing the electric field, the entropy values of the ring-shaped particle assembly first decreased to a minimum and then rapidly increased. The reduction in entropy at this minimum (19.26% from randomly distributed particles, see **Supporting Information – Sec. III**), stood for the highest degree of order in the ring-shaped particle assembly seen for $20\ \mu m$ particles using a $LG_{14}$ laser. As evident in the particle density map at $E_{max} = 3825\ V/m$ (Figure 5C, *green map* inset), 70.1% of the particles gathered around the "forbidden zone" which led to this high degree of order (for more detailed particle maps, see Figure S2). No matter the bin size used for the calculation of entropy, the highest degree of order in the ring-shaped particle assembly was always achieved at $E_{max} = 3825\ V/m$ (Figure S3A). Furthermore, with the smaller $5\ \mu m$ particle size (*red line*, Figure 5C), a minimum entropy value in the ring-shaped pattern was achieved at even lower values of electric field when compared to the $20\ \mu m$ particle ($E_{max} = 2130\ V/m$, Figure 5C and Figure S3B). It is expected that a further reduction in the electric field for the $5\ \mu m$ particle size seen in Figure 5C would further reduce the entropy value (for more detailed particle maps, see Figure S4). This is because the competition between the drag and gravitational forces on the particles, as seen in Figure 4, ultimately dictates the degree of order of the ring-shaped pattern. Since the ratio between the drag forces and the gravitational force is proportional to $d_p^{-2}$, this ratio will increase significantly as the particle size is reduced. Therefore, to improve the entropy of patterns produced with smaller particle sizes, the magnitude of this drag force must be reduced. This can be simply done by lowering the electric field intensity as corroborate in Figure 5C when comparing this effect on both the $20\ \mu m$ and $5\ \mu m$ particle sizes. To further investigate the influence of the drag coefficient on particle assembly, a comparative study between the drag coefficient derived from the Schiller-Naumann model were performed. For both the Stokes and Schiller-Naumann models, a ring-shaped particle pattern was observed, and the trend of entropy change with the electric field intensity was consistent. The only difference between the two models lies in the $E_{max}$ corresponding to the entropy minima, which was higher in the case of the Schiller-Naumann model ($E_{max} = 3950\ V/m$, Figure S5) compared to Stokes law

($E_{max} = 3825\ V/m$) for 20-$\mu m$ particles.

Moreover, it is not only possible to control the extent of the "forbidden zone" through careful choice of the radial and azimuthal index numbers of the $LG_{pl}$ laser, but it is also possible to control the particle dispersion-width of the ring-shaped assembly by tuning the velocity of the fluid with the electric field of the laser. Here the particle dispersion-width is qualitatively defined as the radial distance from the boundary of the "forbidden zone" to the boundary of the liquid-gallium domain wall. The velocity profiles of the fluid as a function of the $LG_{14}$ laser electric field intensity at the liquid-solid interface is shown in Figure 5D. At the boundary of the "forbidden zone" (Figure 5D, *shaded grey region*), the velocity of the fluid dropped to zero. Outside of the "forbidden zone", undulated fluid velocity profiles can be seen. As the electric field intensity of the laser increases, the absolute value for the fluid velocity also increases. This also shifts the fluid velocity maxima away from the center of liquid-gallium (Figure 5D, *dashed black lines*). In this case, the drag forces on the particles would also increase with electric field, making them harder to assemble around the boundary of the "forbidden zone". Consequently, particles would be able to assemble out of the liquid-gallium before reaching the boundary of the "forbidden zone". This would result in a larger dispersion-width in the ring-shaped particle assembly. However, if the fluid velocity becomes too high, the drag forces would dominate the motion of the particles, where particles will prefer to circulate within the convective flow pattern of the fluid at longer time scales, making them more likely to strike the sidewalls. Under this case, particles will gather around the periphery of the liquid-gallium domain wall (Figure S4H). Therefore, to obtain a tighter ring dispersion-width, the intensity of the electric field should be kept at a moderate value to maintain a moderate ratio between the drag and gravitational forces as seen in Figure 4A.

**CONCLUSION**

In summary, light induced surface tension gradients have proven useful in not only controlling fluid flow of opaque liquids, but also directing the trajectory of dispersed particles within the bulk of liquid gallium, which is commonly dictated by Brownian forces. Overcoming Brownian is imperative to realize the controllable movement of solutes within the bulk of liquids, and thus to achieve a predictable spatial assembly with a high degree of order, complementary to what has been realized in particle assembly out of ferrofluids using magnetic fields[23]. Our results highlight that assemblies with different degrees of order can be formed by engineering Marangoni flow in liquid-gallium using $LG_{pl}$ lasers, which represent a tunable approach to design particle assemblies out of liquid metals. We found that the synergy between the drag forces on the particle from convective flow of the fluid and the effective gravitational force are responsible for limiting the influence of Brownian forces on the particle's motion in the fluid, therefore resulting in the formation of unique ring-shaped particle assemblies at the liquid-solid interface with a high degree of tunability. Careful control over the parameters of the $LG_{pl}$ laser (i.e., laser mode, spot size $w_o$, and intensity of the electric field $E_{max}$ etc.) can tune the temperature and fluid dynamics of the liquid-gallium as well as the balance of forces on the particle, which in turns can tune the structure of the ring-shaped particle pattern between the "coffee ring" and "reverse coffee ring" effects. This is a result of prominent "central vortices" within the convective flow of the bulk liquid that creates a "forbidden zone" of assembly at the liquid-solid interface. A striking example of this is when 70.1% of randomly dispersed tungsten particles in liquid-gallium under the $LG_{14}$ laser mode ($d_p = 20\ \mu m$, $w_o = 125\ \mu m$ and $E_{max} = 3825\ V/m$) gathered around the periphery of the "forbidden zone" at the liquid-solid interface, forming a $100\ \mu m$ wide ring-shaped particle assembly that was smaller than the beam size of the laser itself.

Although this study has focused on the interaction of lasers with liquid-gallium, this approach can be extended to other liquid metals, such as mercury (Hg), indium (In), tin (Sn), etc., or eutectic metal alloy systems, such as Galinstan, field's metal, etc. The variety of liquid metals that are available with different melting points, viscosities, solubilities and reactivities will further expand the design space for creating complex material designs from a variety of elements and particle structures dissolved and/or dispersed in these fluids across different length scales from the bottom-up. Additionally, the use of a common laser as a heating source to develop Marangoni flow in liquid metals opens the possibility of achieving rapid spatial switching in the design of the temperature gradients on the liquid metal, and therefore the surface tension gradients. The modularity of this fabrication process could potentially allow for an on demand spatial control over the precipitation of solutes out of the liquid metal, such as that seen in crystal growth processes of nanostructures, like the VLS mechanism [24]. This can be done by altering the input of the laser beam shape and profile of the electric field using spatial light modulators and other appropriate optical components.

**METHODS**

**Framework for coupling the Marangoni effect and fluid flow:**

In this study, liquid-gallium was placed with cylindrical boundary conditions to simplify the model and avoid the influence of surface curvature on surface tension at the two-phase interface. The steady-state Marangoni flow pattern in liquid-gallium induced by the Laguerre-Gaussian laser mode was simulated by coupling the laminar flow, heat transfer in the fluid and electromagnetic wave using a combination of finite element method solver packages (COMSOL Multiphysics®). Herein, the laminar flow in liquid-gallium was governed by the Navier-Stokes equation:

$$\rho_f \vec{v} \cdot \nabla \vec{v} = -\nabla P + \nabla \cdot \left[ \mu(\nabla \vec{v} + (\nabla \vec{v})^T) - \frac{2}{3}\mu(\nabla \cdot \vec{v})\vec{I} \right] + \rho_f \vec{g}$$

where $\rho_f$ is the density of liquid-gallium, $\vec{v}$ is the fluid velocity, $P$ is the pressure in liquid-gallium, $\mu$ is the viscosity of liquid-gallium, and $\vec{g}$ is the gravitational acceleration constant. The heat transfer in the fluid was coupled to the movement of the fluid by combing $\vec{v}$ and $P$. The heat transfer in liquid-gallium was described with the conservation of energy given by:

$$\rho_f C_P \vec{v} \cdot \nabla T + \nabla \cdot \vec{q} = Q_e + Q_P + Q_{vd}$$

$$\vec{q} = -k_T \nabla T$$

where $C_P$ is the heat capacity and $k_T$ is the thermal conductivity of liquid-gallium, respectively. Moreover, the temperature gradient that develops on the surface of liquid-gallium induces sheer stresses at the gas-liquid interface giving rise to Marangoni flow:

$$\left( -P\vec{I} + \mu(\nabla \vec{v} + (\nabla \vec{v})^T) - \frac{2}{3}\mu(\nabla \cdot \vec{v})\vec{I} \right)\vec{n} = k_0 \nabla_t T$$

where $\vec{v}$ is the velocity of the fluid and $k_0$ is the temperature coefficient of surface tension of the fluid. In the case of liquid-gallium, this $k_0$ is negative.

**Laser heating setup and conditions:**

The major heating source was by electromagnetic heating which is given by:

$$Q_e = \frac{1}{2} Re(\vec{J} \cdot \vec{E^*}) = \frac{1}{2} Re(\sigma_f \vec{E} \cdot \vec{E^*})$$

where $\sigma_f$ is the electrical conductivity and $J$ is the current density of the liquid-gallium. In our models, the electromagnetic sources are three different Laguerre-Gaussian laser modes (i.e., $LG_{00}$, $LG_{04}$, and $LG_{14}$ laser modes, propagating in the z-direction) which are described by:

$$E_{p,l}(r,z) = E_{0_{p,l}}(r,z) \frac{1}{\sqrt{1 + \frac{z^2}{z_0^2}}} \exp\left[ -\frac{r^2}{w_0^2(1 + \frac{z^2}{z_0^2})} - ikz - ik\frac{r^2}{2z\left(1 + \frac{z_0^2}{z^2}\right)} + i arctan\left(\frac{z}{z_0}\right) \right]$$

where $z_0$ is the Rayleigh range of the laser and $E_{0_{p,l}}(r,z)$ is the amplitude for the electric field distribution of the $LG_{pl}$ laser mode.

For $LG_{00}$ laser mode:

$$E_{0_{0,0}}(r,z) = E_0$$

For $LG_{04}$ laser mode:

$$E_{0_{0,4}}(r,z) = E_0 \times \left[\frac{\sqrt{2}r}{w(z)}\right]^4 \times \exp\left[4 \times \left(i\phi + i arctan\left(\frac{z}{z_0}\right)\right)\right]$$

For $LG_{14}$ laser mode:

$$E_{0_{1,4}}(r,z) = E_0 \times \left(\frac{\sqrt{2}r}{w(z)}\right)^4 \times [5 - \frac{2r^2}{w^2(z)}] \times \exp\left[4 \times i\phi + 6 \times i arctan\left(\frac{z}{z_0}\right)\right]$$

$$w(z) = w_0 \sqrt{1 + \frac{z^2}{z_0^2}}$$

$$\phi = \text{atan2}(\frac{y}{x})$$

Furthermore, the work done by pressure change in liquid-gallium ($Q_P$) is given by:

$$Q_P = \alpha_P T \vec{v} \cdot \nabla P$$

where, $\alpha_P$ is the coefficient of thermal expansion of liquid-gallium. The viscous dissipation in liquid-gallium ($Q_{vd}$) is given by:

$$Q_{vd} = \tau : \nabla \vec{v}$$

where $\tau$ is the viscous tensor. At the boundary of liquid-gallium, radiation heating from the liquid-gallium to ambient (air) was taken into consideration. The net outward radiative flux ($q_{r,net}$) is given by:

$$q_{r,net} = \varepsilon(e_b(T) - G) = \varepsilon(n^2 \sigma T^4 - G)$$

$$\vec{n} \cdot \vec{q} = q_{r,net}$$

where $\varepsilon$ and $n$ are the emissivity and refractive index of liquid-gallium at $645\ nm$, respectively. $\sigma$ is the Stefan–Boltzmann constant and $G$ is the total incoming radiative flux. $\vec{n}$ is the unit normal vector.

**Particle trajectory simulation setup:**

Particle trajectories were simulated under steady state fluid flow. A random distribution of 2000 neutrally charged tungsten particles were consistently released in the liquid-gallium for fair comparison. The particle trajectory was simulated based on Newton's second law:

$$\frac{d(m_P \vec{u})}{dt} = \vec{F_D} + \vec{F_B} + \vec{F_G} + \vec{F_{Bouyant}}$$

where $\vec{F_D}$ is the drag force from the fluid flow, $\vec{F_B}$ is the Brownian force, $\vec{F_G}$ is the gravitational force, and $\vec{F_{Bouyant}}$ is the buoyant force. To guarantee the accuracy of the simulated particle trajectories, the time step was set to the characteristic time of the particle ($\tau_P$), which is given by:

$$\tau_P = \frac{\rho_p d_p^2}{18\mu}$$

The characteristic time for the $20\ \mu m$ and $5\ \mu m$ tungsten particles was $6.72 \times 10^{-4}\ s$ and $4.2 \times 10^{-5}\ s$, respectively. Gravitational, buoyant, Brownian, and drag forces were applied to the particles. For the boundary conditions, all particles bounced back elastically, except at the liquid-solid interface where assembly of particles can occur (bottom side of liquid-gallium). For the liquid-solid interface, a sticking probability of 50% was included in our models, otherwise the particle would be able to bounce back into the fluid elastically. To simplify the calculations, the heat transfer between particles and liquid-gallium was neglected due to the low concentration of particles in the fluid (<10$^{-15}\ mol/L$). Also, the particles are not interacting with each other. More details on the model settings are included in the **Supporting Information**.

## ASSOCIATED CONTENT

### Supporting Information

Additional information including details of physics setting for the particle trajectory simulation, calculation of mean-squared displacement for particle trajectory, entropy calculation of the particle assemblies, probability of the particle-sidewall interactions, and particle patterns under different maximum intensity value of the electric field.

## AUTHOR INFORMATION

### Corresponding Author

**Zakaria Y. Al Balushi -** Department of Materials Science and Engineering, University of California, Berkeley, Berkeley, CA 94720, USA; Materials Sciences Division, Lawrence Berkeley National Laboratory, Berkeley, CA 94720, USA; Email: albalushi@berkeley.edu.

**Author**

**Jiayun Liang -** Department of Materials Science and Engineering, University of California, Berkeley, Berkeley, CA 94720, USA

**Author contributions**

Both authors contributed extensively to this work, including in the analysis and interpretation of the results. J.L. performed all the calculations. Z.A. conceived the idea and supervised the project. Both authors discussed, wrote, revised, and approved the manuscript.

**Notes**

Authors declare that they have no competing interests.

**ACKNOWLEDGEMENT**

We thank Jiawei Wan for helpful discussions on the calculation of entropy for particle patterns and Dr. Xiaotian Zhang on simulation troubleshooting. All the simulations were performed at the Molecular Graphics and Computation Facility (MGCF) in University of California, Berkeley. This work was supported by the Laboratory Directed Research and Development Program of Lawrence Berkeley National Laboratory under U.S. Department of Energy Contract No. DE-AC02-05CH11231.

**DATA AVAILABILITY STATEMENT**

All data needed to evaluate the conclusions in the paper are present in the paper and/or the Supporting Information.

# Supporting information

# Light Induced Surface Tension Gradients for Hierarchical Assembly of Particles from Liquid Metals

Jiayun Liang[1], Zakaria Y. Al Balushi[1,2]

[1]Department of Materials Science and Engineering, University of California, Berkeley; Berkeley, CA 94720, USA.

[2]Materials Sciences Division, Lawrence Berkeley National Laboratory, Berkeley, CA 94720, USA.

*Corresponding author, e-mail: albalushi@berkeley.edu

**This PDF file includes:**

Supporting Text

- Sec. I: Physics setting for the particle trajectory simulation
- Sec. II: Mean-squared displacement for the particle trajectory
- Sec. III: Entropy calculation of the particle assemblies
- Sec. IV: Probability of particle-sidewall interaction

Supporting Tables

- Table S1: Summary of elements that do not form compounds with liquid gallium stable above the energy hull cutoff
- Table S2: Parameters for finite element model

Supporting Figures

- Figure S1. XY cross-section of the particle assembly with the sticking coefficient of the sidewalls set to unity ($LG_{14}$ laser mode and $w_0 = 125 \ \mu m$)
- Figure S2. Particle patterns under different maximum intensity value of the electric field ($E_{max}$) for particles with a diameter of $20 \ \mu m$ ($LG_{14}$ laser mode and $w_0 = 125 \ \mu m$)
- Figure S3. Entropy of the particle pattern under different maximum intensity value of the electric field ($E_{max}$) calculated with different bin sizes ($LG_{14}$ laser mode and $w_0 = 125 \ \mu m$)
- Figure S4. Particle patterns under different maximum intensity value of the electric field ($E_{max}$) for particles with a diameter of $5 \ \mu m$ ($LG_{14}$ laser mode and $w_0 = 125 \ \mu m$)
- Figure S5. Entropy of particle patterns under different $E_{max}$ values ($LG_{14}$ laser mode and $w_0 = 125 \ \mu m$) with drag coefficient from Schiller-Naumann model.
- Figure S6. Statistical analysis of the entropy calculations

# SUPPORTING TEXT

## Sec. I: Physics setting for the particle trajectory simulation

The steady-state Marangoni flow pattern in liquid gallium induced by the Laguerre-Gaussian laser mode was simulated by coupling the laminar flow, the heat transfer in the fluid and the electromagnetic wave using a finite element method solver COMSOL Multiphysics®. The size of liquid gallium was $1000\ \mu m$ in diameter and $50\ \mu m$ in height. To guarantee the accuracy of the simulation results, an extremely fine and symmetric mesh containing 173,888 triangle elements was built.

The particle trajectory was simulated based on Newton's second law:

$$\frac{d(m_P \vec{u})}{dt} = \vec{F_D} + \vec{F_B} + \vec{F_G} + \vec{F_{Bouyant}} \tag{1}$$

where $\vec{F_D}$ is the drag force from the fluid flow, $\vec{F_B}$ is the Brownian force, $\vec{F_G}$ is the gravitational force, and $\vec{F_{Bouyant}}$ is the buoyant force. Here, the drag force from the fluid flow ($\vec{F_D}$) is given by:

$$\vec{F_D} = \frac{m_P(\vec{u} - \vec{v})}{\tau_P} \tag{2}$$

where $\vec{u}$ is the velocity of particle, $\vec{v}$ is the velocity of fluid flow, and $\tau_P$ is the characteristic time for the particle. The amplitude of the Brownian force ($\vec{F_B}$) is given by:

$$\vec{F_B} = \zeta \sqrt{\frac{6\pi k_B \mu T d_p}{\tau_P}} \tag{3}$$

where $k_B$ is the Boltzmann constant and $\zeta$ is a vector of independent, normally distributed random numbers with zero mean and unit standard variation as indicated in COMSOL. For the sticking probability of the particles at the liquid-solid interface was set to 50% (bottom side of liquid gallium). When particle stick on the boundary:

$$\vec{u} = \vec{0} \tag{4}$$

otherwise, the particle would bounce elastically back into the fluid:

$$\vec{u} = \vec{u_c} - 2(\vec{n} \cdot \vec{u_c})\vec{n} \tag{5}$$

where $\vec{u_c}$ is the particle velocity when the particle approaches the boundary of the liquid gallium domain. At all other boundaries of liquid gallium, the particles would bounce back elastically.

## Sec. II: Mean-squared displacement for the particle trajectory

The mean-squared displacement ($MSD\ (\tau_P)$) for the particle trajectory is given by:

$$MSD(\tau_P) = <[\vec{r}(t + \tau_P) - \vec{r}(t)]^2> \tag{6}$$

where $\vec{r}$ is the position of the particle at time $t$ and $\tau_P$ is the time step for the simulation. For all the datasets presented in Figure 2 of the main text, the time step was set as the characteristic time of the particle, i.e., for $20\ \mu m$ tungsten particles, $\tau_P$ was $6.72 \times 10^{-4}\ s$.

**Sec. III: Entropy calculation of the particle assemblies**

In all results, the number of particles that were released into liquid gallium was 2000. We then used entropy to quantify the degree of order of the particle pattern assembly at the liquid-solid interface. To calculate the entropy for different particle assemblies, we equally divide the particle pattern into $(\frac{1000}{n})^2$ square lattices (i.e., "bin") of size $n$ and counted the number of particles in each bin. Here, the entropy for the patterns is defined as:

$$S = -\sum_{i=1}^{\left(\frac{1000}{n}\right)^2} \frac{N_i}{2000} \ln \frac{N_i}{2000} \tag{7}$$

where $N_i$ is the number of particles in the $i^{th}$ bin.

To test the reproducibility of our calculations, we repeated each simulation 10 times and compared the entropy for the particle assembly obtained in different runs. A series of bin sizes were utilized for entropy calculation to avoid the impact of the bin size on the entropy results. As shown in Figure S6, no matter the size of the bin utilized, the relative entropy difference was less than 0.8% and the mean squared error was less than 0.6%, supporting the reproducibility of our calculations.

In addition, to quantify the enhancement of the degree of the order, we set the entropy for a completely random particle assembly composed of 2000 tungsten particles as the reference value ($S_{ref}$). In a completely random particle assembly, the possibility for a particle to assemble at different positions is the same. Therefore, with a $50\ \mu m$ bin size, the entropy for a completely random particle assembly is given by:

$$S_{ref} = -\sum_{i=1}^{400} \frac{5}{2000} \ln \frac{5}{2000} = 5.9915 \tag{8}$$

Therefore, the reduction in entropy ($\Delta S$) is given by:

$$\Delta S = \frac{S_{ref} - S_{particle\ assembly}}{S_{ref}} \tag{9}$$

where $S_{particle\ assembly}$ is the entropy for the specific particle assembly of interest. $S_{ref}$ and $S_{particle\ assembly}$ are calculated with the same bin size.

**Sec. IV: Probability of particle-sidewall interaction**

A particle can only interact with the sidewall when the distance between the particle and the sidewall is no more than the radius of the particle itself. Based on this assumption, the probability for the particle-sidewall interaction ($P_0$) with different particle size can be given by:

$$P_0 = \frac{\pi\left(d_f^2 - (d_f - d_p)^2\right)}{\pi d_f^2} = \frac{2d_f d_p - d_P^2}{d_f^2} \qquad (10)$$

where $d_f$ is the diameter of the liquid gallium ($1000\ \mu m$ in our model). The probability of the particle-sidewall interaction for $20\ \mu m$ diameter particles is 3.96%, while for $5\ \mu m$ diameter particles, the probability drops down to 0.998%. Therefore, as shown in Figure S1, for $5\ \mu m$ diameter particles, less particles stick to the sidewalls (Figure S1B, sticking probability = unity), compared to the pattern produced from $20\ \mu m$ diameter particles (Figure S1A, sticking probability = unity).

**SUPPORTING TABLES**

**Table S1.** Summary of elements that do not form compounds with liquid gallium stable above the energy hull cutoff [1]

| H | He | Be |
|---|---|---|
| B | C | Ne |
| Al | Si | Ar |
| Zn | Ge | Kr |
| Cd | In | Sn |
| Xe | W | Hg |
| Tl | Pb | Bi |

**Table S2.** Parameters for finite element model [2]

| Parameter | Physical Meaning | Value |
|---|---|---|
| $k_0$ | Temperature coefficient of surface tension for liquid gallium | $-6.6 \times 10^{-5}\ N/(m*K)$ |
| $C_P$ | Heat capacity for liquid gallium | $381.072862\ J/(kg*K)$ |
| $\rho$ | Density of liquid gallium | $5592\ kg/m^3$ |
| $\rho_p$ | Density of tungsten particle | $1930\ kg/m^3$ |
| $k_T$ | Thermal conductivity of liquid gallium | $8.0844\ W/(m*K)$ |
| $\alpha$ | Thermal expansion of liquid gallium | 3.0762 |
| $\mu$ | Dynamic viscosity of liquid gallium | $0.64288\ mPa*s$ |
| $\sigma_f$ | Electric conductivity of liquid gallium | $7.1\times 10^6\ S/m$ |

| $n$ | Real part of refractive index for liquid gallium at 645nm | $n = 1.57434 + 6.38 * 10^{-4} * (T - 273.15)$ |
|---|---|---|
| $k$ | Imaginary part of refractive index for liquid gallium at 645nm | $k = 7.48476 + 2.46 * 10^{-3} * (T - 273.15) + 2.18 * 10^{-6} * (T - 273.15)^2$ |
| $\varepsilon$ | Emissivity of liquid gallium at 645 nm | $\varepsilon = 0.0771 + 6.852 * 10^{-5} * T$ |

**SUPPORTING FIGURES**

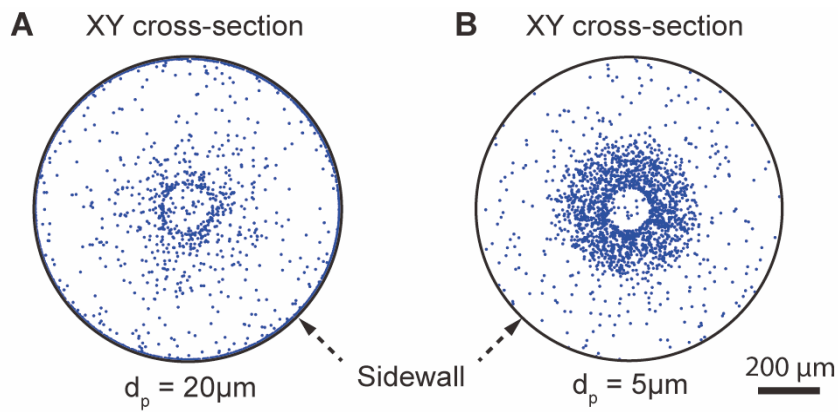

**Figure S1. XY cross-section of the particle assembly with the sticking coefficient of the sidewalls set to unity ($LG_{14}$ laser mode and $w_0 = 125\ \mu m$).** The diameter of the particles ($d_p$) are: **(A)** 20 $\mu m$, and **(B)** 5 $\mu m$, respectively. The sidewall is marked with *solid black line* and highlighted with a *dashed black arrow*. Scale bar for the particle pattern is 200 $\mu m$.

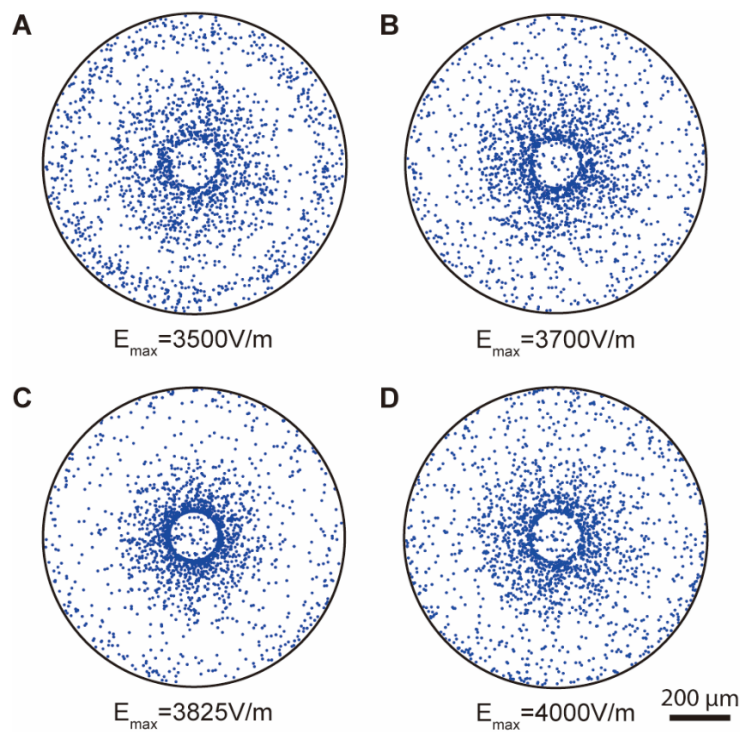

**Figure S2. Particle patterns under different maximum intensity value of the electric field ($E_{max}$) for particles with a diameter of 20 $\mu m$ ($LG_{14}$ laser mode and $w_0 = 125\ \mu m$). (A)** $E_{max} = 3500\ V/m$, **(B)** $E_{max} = 3700\ V/m$, **(C)** $E_{max} = 3825\ V/m$, and **(D)** $E_{max} = 4000\ V/m$. Scale bar for the particle pattern is 200 $\mu m$.

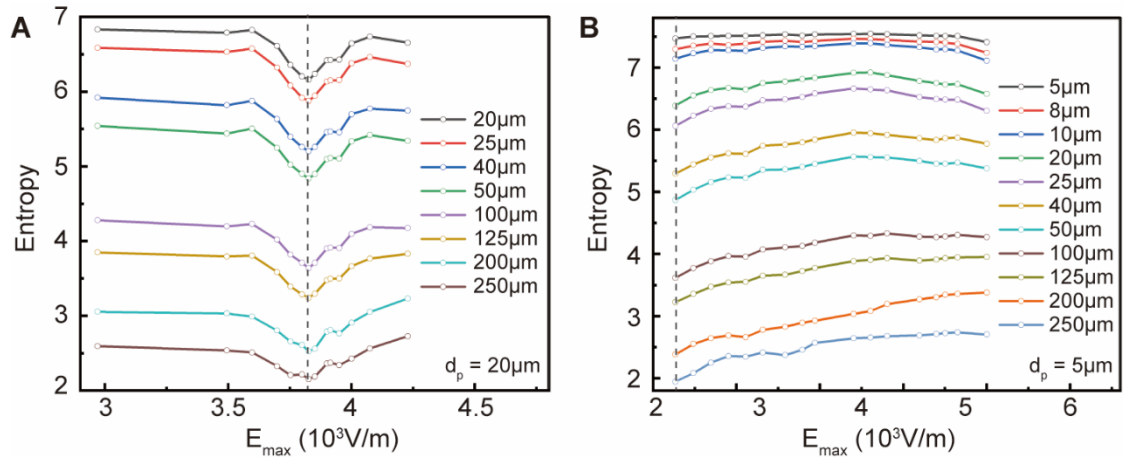

**Figure S3. Entropy of the particle pattern under different maximum intensity value of the electric field ($E_{max}$) calculated with different bin sizes ($LG_{14}$ laser mode and $w_0 = 125\ \mu m$).** The diameter of the particles ($d_p$) are: **(A)** 20 $\mu m$, and **(B)** 5 $\mu m$, respectively.

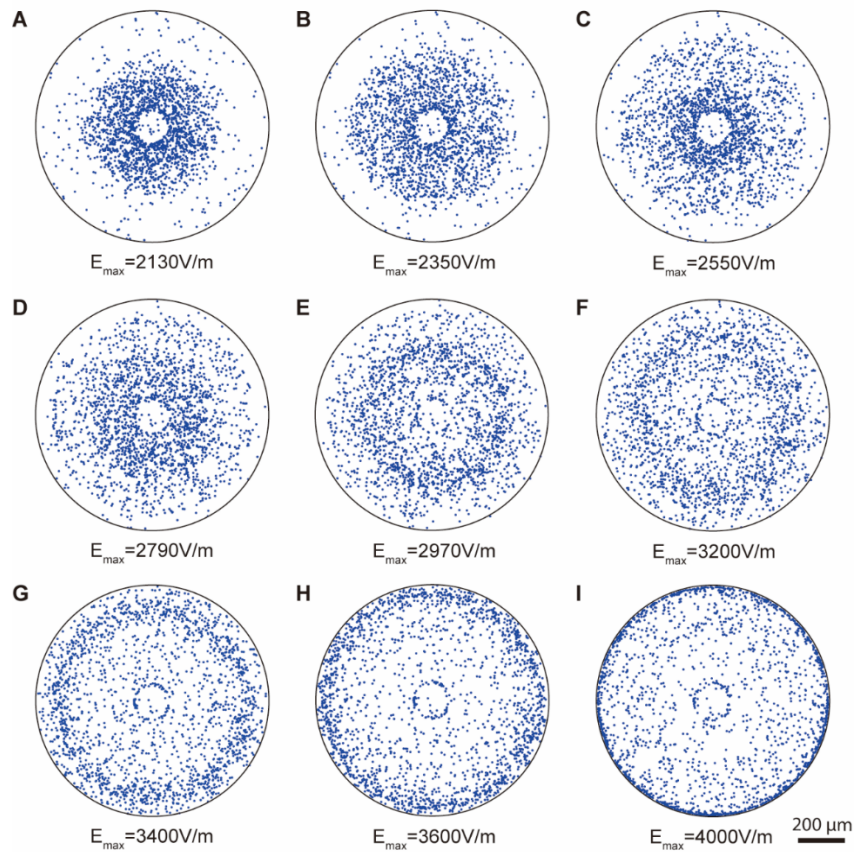

**Figure S4. Particle patterns under different maximum intensity value of the electric field ($E_{max}$) for particles with a diameter of 5 $\mu m$ ($LG_{14}$ laser mode and $w_0 = 125\ \mu m$). (A)** $E_{max} = 3500\ V/m$, **(B)** $E_{max} = 2130\ V/m$, **(C)** $E_{max} = 2350\ V/m$,

**(D)** $E_{max} = 2550\ V/m$, **(E)** $E_{max} = 2790\ V/m$, **(F)** $E_{max} = 2970\ V/m$, **(G)** $E_{max} = 3400\ V/m$, **(H)** $E_{max} = 3600\ V/m$, and **(I)** $E_{max} = 4000\ V/m$. Scale bar for the particle pattern is $200\ \mu m$.

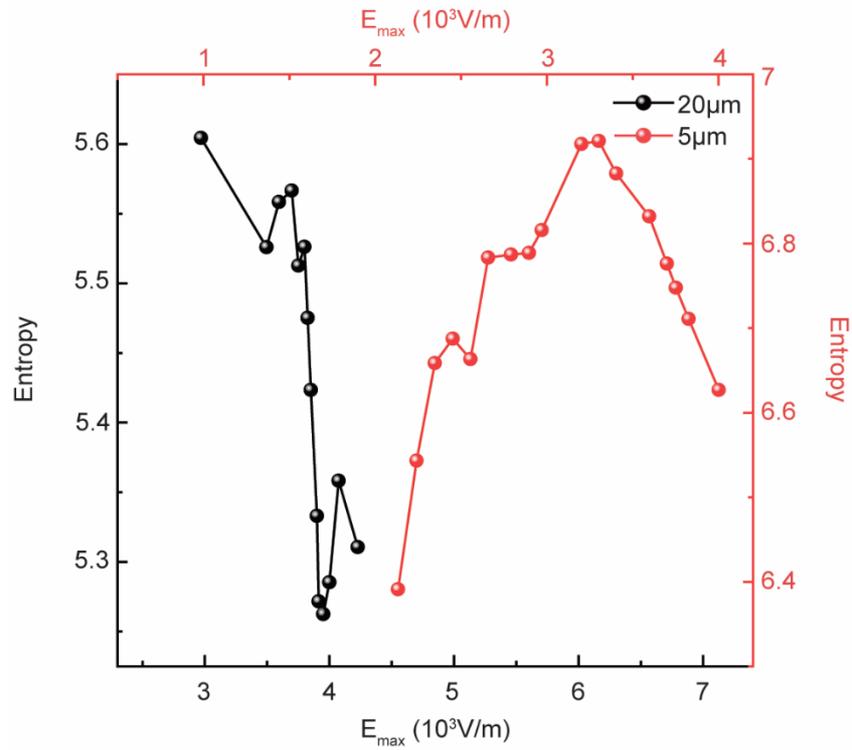

**Figure S5. Entropy of particle patterns under different $E_{max}$ values ($LG_{14}$ laser mode and $w_0 = 125\ \mu m$) with drag coefficient from Schiller-Naumann model.** The bin size for the calculation of entropy is $50\ \mu m$ for $20\ \mu m$ diameter particles (*solid black line*), and the bin size is $20\ \mu m$ for $5\ \mu m$ diameter particles (*solid red line*).

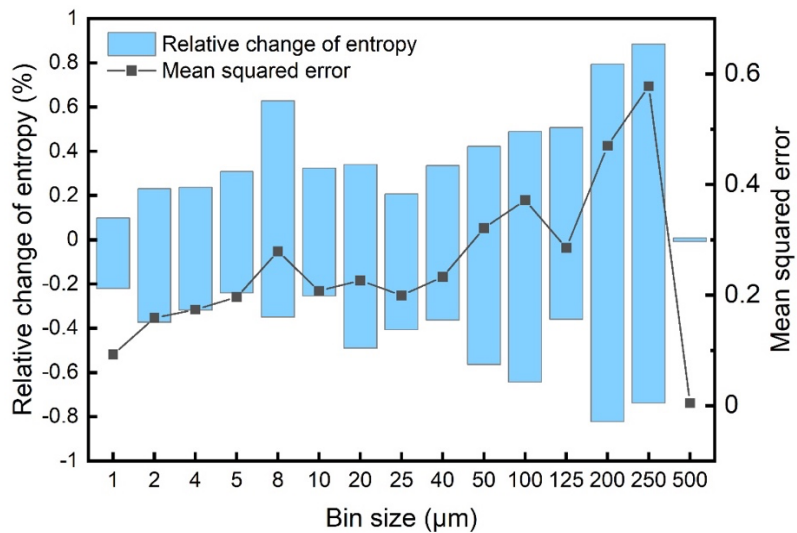

**Figure S6. Statistical analysis of the entropy calculations.** Relative change of entropy (left y-axis, *blue box*) and mean squared error of entropy (right y-axis, black solid line) for the particle assembly obtained in 10 repeated runs.

## SUPPORTING REFERENCES